# Broadband probing magnetization dynamics of the coupled vortex state Permalloy layers in nanopillars

A. A. Awad[1], A. Lara[1], V. Metlushko[2], K.Y. Guslienko[3,4], and F. G. Aliev[1*]

[1] *Dpto. Física de la Materia Condensada C-III, Universidad Autónoma de Madrid, 28049 Madrid Spain*

[2] *Dept. Electrical and Computer Engineering, University of Illinois, Chicago, 60607 IL, USA*

[3] *Dpto. Física de Materiales, Universidad del País Vasco, 20018 San Sebastian, Spain*

[4] *IKERBASQUE, The Basque Foundation for Science, 48011 Bilbao, Spain*

Broadband high-frequency magnetization response of the coupled vortex state magnetic dots in layered nanopillars was explored as a function of in-plane magnetic field and interlayer separation. For dipolarly coupled circular Py(25nm)/Cu(20nm)/Py(25nm) nanopilars of 600 nm diameter a small in-plane field splits the eigenfrequencies of azimuthal spin wave modes inducing an abrupt transition between in-phase) and out-of-phase kinds of the low-lying coupled spin wave modes. The critical field for this splitting is mainly determined by antiparallel chiralities of the vortices in the layers. Qualitatively similar (although more gradual) changes occur also in the exchange coupled Py(25nm)/Cu(1nm)/Py(25nm) tri-layer nanopillars. These findings are in qualitative agreement with micromagnetic dynamic simulations.

(*) farkhad.aliev@uam.es

Intensive experimental and theoretical investigations of the magnetic vortex dynamics in small particles (dots) during the last decade[1] opened new ways to process information stored in the vortex core[2], vortex chirality[3,4] and to transmit it between the coupled dots using vortex dynamics.[5-8] Actually, circular soft magnetic dots in the vortex ground state are the main elements of many proposed novel spintronic devices[9-11], capable of fascinating spin-based electronics applications, from extremely sensitive magnetic field sensors, to spin polarized current-tunable microwave vortex nano-oscillators[10] and vortex MRAM[11].

Layered magnetic nanopillars consisting of ferromagnetic (F) and nonmagnetic layers (N) are becoming the main "building blocks" in spintronics.[12] However, the bulk of the current knowledge on magnetization dynamics of the vortex state dots (so called gyrotropic and spin wave modes of different symmetry) is related to a single vortex state dot, or laterally coupled vortex dots[6-8]. As to the vertically coupled magnetic vortices in F/N/F nanopillars, the scarce knowledge on dynamics is mainly restricted to investigation of the interlayer coupling effect on the vortex low frequency gyrotropic modes[10,13] and a little is known on the influence of character of the coupling[14] on the excited high frequency spin wave modes[15]. Besides, the major part of the previous studies of the vertically coupled vortex state dots in nanopillars has been done in asymmetric conditions by using the dots made of different materials (typically Co and Py)[14,16] or having different thickness.[17] Using symmetric (the same thickness and material) circular ferromagnetic layers having degenerated eigenfrequencies of the isolated layers would give an unique opportunity to investigate the influence of the strength and character of interlayer coupling (i.e., interlayer exchange vs. dipolar) and relative vortex chiralities on spin wave dynamics in vortex-vortex coupled system without influence of additional factors such as Co anisotropy. The experiments[18] showed that some conditions should be satisfied to stabilize a remanent double vortex state in F/N/F stack. If the F-layer thicknesses and radii are relatively small, the dots are in in-plane single domain state and only edge localized modes can be detected in the low frequency part of the spin excitation spectra[15].

Here we explore, experimentally, and by simulations and analytical estimations, the dynamic properties of the vertically coupled permalloy (Py) dots with two spacer thicknesses of 0.9 and 20 nm in Py/Cu/Py circular nanopillars of relatively large radius $R$=300 nm. We investigate high-frequency response by applying a bias in-plane field and compare the results with the spin wave spectra of uncoupled dots[19,20].

Two set of samples in the form of square arrays of tri-layer Py(25nm)/Cu(d)/Py(25nm) circular dots were fabricated by a combination of e-beam lithography and lift-off techniques on a standard Si(100) substrate as explained elsewhere[21]. The samples have Py layers thickness of $L$=25 nm, diameter $2R$=600 nm and large interdot center-to-center distances of 1000 nm to minimize the dipolar lateral coupling. The thickness of the Cu spacer for the first type of samples is $d$=0.9 nm (we call it as IEC-interlayer exchange coupled tri-layer), while the second type of samples has $d$=20 nm (further- DIC, dipolar interlayer coupled tri-layer).

The spin wave spectra of the nanopillars have been studied at room temperature by broadband vector network analyzer based VNA-FMR technique[20,21] in the range 6-12 GHz. Continuous reference layers have also been studied separately to characterize the interlayer coupling. The data were obtained and analyzed using parallel pumping scheme when both the bias dc field $\mathbf{H}_0 \parallel \mathbf{x}$ and oscillating pumping field $\mathbf{h}$ are in plane and parallel to each other (see Fig. 1). The pumping field $\mathbf{h}$ excites only the spin eigenmodes localized in the areas where the torque $\mathbf{h} \times \mathbf{M}_0 \neq 0$, ($\mathbf{M}_0$ is the static magnetization) and quasi-uniform FMR-mode is suppressed[20]. The hysteresis loops for both IEC and DIC tri-layer nanopillars (not shown) are in qualitative agreement with previous results on hysteresis loops for the vertically coupled circular dots in F/N/F nanopillars with vortices present in both the F-layers at zero field[18].

To further confirm the presence of the double vortex ground state in our samples, we carried out micromagnetic simulations[22]. We characterize the vortex in j-layer ($j$=1,2) by its core polarizations $p_j$ and chirality $C_j$.[1] As we demonstrate below, magnetization dynamics of DIC nanopillars points out (as confirmed by static simulations) the existence of a statistical distribution of antiparallel (APC, $C_1C_2$= -1) and parallel chirality (PC, $C_1C_2$=+1) states of the DIC and IEC dots.

Analysis of the magnetization reversal reveals that the vortex annihilation $H_{an}$ fields are larger for IEC than for DIC nanopillars indicating that the dipolar coupling, present in both IEC and DIC pillars, is strengthened in the IEC dots. In order to investigate in detail the field dependent changes in dynamic response of the coupled dots, we carry out the experiments by measuring averaged broadband signal using multiple scans within the minor loops without annihilating any of the two vortices in the dots. Both DIC and IEC dots reveal clearly strong changes of the spin wave response at low fields (Figs. 1, 2). We concentrate first on our main result: the spin wave excitations for DIC dots (Fig. 1), where frequency doublets were observed in the small field region. These doublets then transform in multiple satellites with increasing bias field above some critical value of about 40 Oe. The bias field induced changes of the eigefrequencies for DIC dots are more abrupt than for the IEC ones.

To identify the observed frequency peaks, we employ standard schematics describing the spin waves (SW) excited in nearly centered vortex state according to number of the nodes in dynamic magnetization observed along the azimuthal $m$ and radial $n$ directions. As the vortex core radius is about 10 nm and the dot radius is 300 nm, the spin wave modes are concentrated mainly outside the core. The SW can be described via dynamic magnetization $\mathbf{m} = (m_\rho, 0, m_z)$ components in the cylindrical coordinates $(\rho, \varphi, z)$, $m_z(\mathbf{\rho},t) = a_{nm}(\rho)\cos(m\varphi - \omega t)$, $m_\rho(\mathbf{\rho},t) = b_{nm}(\rho)\sin(m\varphi - \omega t)$. Accounting that driving magnetic field $\mathbf{h}$ is linearly polarized and oscillates in the dot plane $xOy$, only the azimuthal modes with with $m = \pm 1$ (arbitrary $n$) can be excited. These spin waves correspond to rotating in plane average magnetization $\langle m_x(t)\rangle_V \sim C\langle b\rangle_\rho \sin(\omega t)$, $\langle m_y(t)\rangle_V \sim C\langle b\rangle_\rho m\cos(\omega t)$. So, maximum mode intensity corresponds to minimal number of oscillations of the mode profile $b(\rho)$ along the radial direction, *i.e.*, to the index $n$=0. In zero approximation (no interlayer coupling) the high frequency part of the nanopillar spectrum consist of two peaks – azimuthal SW with $n$=0, $m = \pm 1$. The frequency degeneracy of these SW is removed due to the dynamic vortex-SW interaction resulting in forming of the doublets with the frequency splitting of 1.3 GHz[19,23] ($L$=25 nm, $R$=300 nm). The interlayer coupling energy in the F/N/F stack consists of two parts: exchange (essential only if $d$< 2 nm) and magnetostatic coupling (essential for all $d$). The volume density of the exchange energy can be written as $w_{int}^{ex} = -(J/LM_s^2)\mathbf{m}_1 \cdot \mathbf{m}_2$. The magnetostatic coupling energy density in the main approximation can be written via the F-layer dipole moments $w_{int}^{dip} = (V/(d+L)^3)\langle\mathbf{m}_1\rangle \cdot \langle\mathbf{m}_2\rangle$ placed in the centers of the dots of volume V. We see that the corresponding interaction fields $H_{int}^{ex}$ and $H_{int}^{dip}$ can be added to each other because they follow the same angular dependence $\sim \cos\Theta$, where $\Theta$ is the angle between the averaged layer dynamic magnetizations. Since for Cu the maximum value of $J \sim 0.14$ erg/cm$^2$,[25] estimation shows that the magnetostatic coupling dominates for both the Cu-spacer thicknesses $d$=0.9 and 20 nm explored. The value of $H_{int}^{dip}$ is about 100 Oe in agreement with the estimation from the Py layer hysteresis loops[24], so, the corresponding shifts of the eigenfrequencies of isolated Py layers due to the interlayer coupling should be about of 300 MHz. The dipolar interaction energy is $w_{int}^{dip} \propto \langle\mathbf{m}_1\rangle \cdot \langle\mathbf{m}_2\rangle \propto -C_1 C_2 m_1 m_2$, where $m_1, m_2 = \pm 1$ are the indices of the azimuthal SW forming the doublets in the 1$^{st}$ and 2$^{nd}$ layers. The product $m_1 m_2 = p_1 p_2$ because the sign of $m_j$ is determined by sign of $p_j$ of the j-th layer. The positive (negative) sign of $C_1 C_2 p_1 p_2$ corresponds to effective ferromagnetic (antiferromagnetic) interlayer coupling. Each of the $(0, m_j)$-frequencies splits into the frequencies of in-phase and out-of-phase modes.[24] The in-phase (out-of-phase) mode frequency is lower for ferromagnetic (antiferromagnetic) coupling.

Summarizing, we can say that the eigenfrequency spectrum of F/N/F tri-layer pillars is formed mainly by the intralayer magnetostatic interaction yielding spin wave modes of the different symmetry. Then, the excited azimuthal SW with $m=\pm 1$ are splitted due to interaction with the dynamic vortex cores, and the resulted frequencies are renormalized by relatively weak interlayer coupling. But, nevertheless, this coupling is important because it distinguishes the modes excited in the $C_1C_2$=+/-1 chirality states and leads to formation of in-phase/out-of-phase modes in F-layers and to a strong dependence of the spectra on the external bias fields. We note that an opposite classification of the simulated SW spectra was done for Py(20nm)/Cu(10nm)/Py(10 nm) nanopillars[26] assuming that the interlayer interaction is essentially larger than the azimuthal mode frequency splitting. Only in-phase modes can be excited by uniform **h**, and, therefore, detected by the FMR technique.

Dynamic OOMMF simulations[22] with parallel vortex cores show qualitative agreement with our main experimental observations. The simulations are done by applying a time dependent driving field (Gaussian pulse) with the amplitude of 5 Oe and full width at half maximum (FWHM) of 1 ps to the relaxed state. By performing local Fourier transforms for all the simulation cells and averaging these spectra we can obtain the spin eigenfrequencies. In order to identify the coupled vortex spin wave modes the spatial distribution of the main eigenmodes is obtained by reconstruction of the eigenmode profiles from the local distributions of the phases and amplitudes for every cell for a selected eigenfrequency. Simulations of the ground state at zero field for DIC dots show the lowest azimuthal modes that are splitted due to the dynamical vortex core – SW interaction in each F-layer. Zero field experiments reveal, however, the existence of a pair of the doublets in the frequency range where the lowest azimuthal mode doublet is expected. The further splitting of the $(0, m_j)$ mode frequencies of the doublets is attributed to the coupling of the layers having the antiparallel (APC) or parallel (PC) chiralities in each of the DIC dots forming the tri-layers, as indicated by simulations below (Fig. 1). As the eigenfrequencies of the isolated layers are degenerated, the splitting of the frequencies for different signs of $C_1C_2$ is a result of the interlayer coupling. In agreement with experimental observations, the finite field simulations show an abrupt additional splitting and the relative phase changes in the SW azimuthal modes applying magnetic field exceeding 15 Oe. We need to assume a mixture of the $C_1C_2=+1/-1$ state nanopillars in the measured array to explain the splitting of the modes $(0,\pm 1)$. The assumption about 50% mixture (we have checked with simulations that approximately half of the DIC tri-layers relaxes into PC state, and the other half - into APC state, starting from a random distribution of magnetization) of the APC and PC nanopillars describes qualitatively well the experimental observations with four weakly field dependent spin wave modes at small fields, which are transformed to multiple and strongly field dependent SW frequencies above some critical magnetic field (Fig. 1). The frequencies of nanopillars with $C_1C_2 = -1$ (higher frequency in-phase mode) experience splitting in smaller bias fields than one for nanopillars with $C_1C_2 = +1$ (lower frequency in-phase mode). I.e., the SW modes of nanopillars with $C_1C_2 = +1$ are less sensitive to change of the bias field. Figure 2 compares experiments and simulations for the low field response in IEC dots for 70% APC - 30% PC distributed chiralities. Similar, but more gradual (than observed in DIC dots) changes are seen in the spin wave eigenmode frequencies varying the bias field.

Figure 3 compares simulated spatial distributions of the vortex dynamic magnetization in the DIC dots with PC (part b) and APC chiralities (part a), presented as $\Delta M_x/M_s$, which suffers strong changes above 15 Oe. As shown in Fig.3a, at zero field, the main two modes occur at 8.13 GHz (counter-clockwise (CCW) motion, $m=+1$) and at 10.15 GHz (clockwise (CW) motion, $m=-1$). Being of the acoustic type (i.e., the two layers response is nearly in-phase, and the frequency of in-phase APC mode (a) is higher than the frequency of in-phase PC mode (b)) at the zero bias (H=0), the modes have out-of-phase character when a small external magnetic field above 15 Oe is applied (Fig. 3a). We believe that this occurs because at fields different from zero, each vortex is pushed towards a different part of the dot, and differences of phase appear as a result of the changes in the symmetry of the ground state with respect to the dot centre. At fields higher than 20 Oe the first mode splits into other two, at 7.97 (CCW) and 8.35 GHz (CCW), while the upper mode also splits into other two modes with frequencies 9.99 (CW) and 10.26 GHz (CW). No extra splitting occurs for the PC nanopillar (Fig. 3b), and the dynamical $\Delta M_x/M_s$ distribution changes slowly increasing the bias field. The zero field splitting of the APC and PC in-phase modes in Fig. 3 (0.55 GHz and 0.41 GHz for the low/high doublet frequency) is a direct measure of the interlayer coupling.

To summarize, magnetization dynamics in coupled vortex dots in Py/Cu/Py tri-layer nanopillars applying in plane magnetic field reveals substantial differences in respect to a single dot response. For APC dipolar coupled dots external in plane magnetic field exceeding small (few tens of Oe) critical value leads to excitation of multiple spin wave modes, characterized not only by their strong field dependence, but also their differences in relative phase of the dynamic response of the layers. Knowledge of the spin eigenmodes of coupled tri-layer pillars is of considerable importance for understanding of thermally induced noise in nanopillar based read heads and spin wave modes excited in spin torque and magnonic devices.

Authors acknowledge support from Spanish MINECO (MAT2009-10139; CSD2007-00010) and CM (P2009/MAT-1726). K.G. acknowledges support by IKERBASQUE and partially support by MEC Grants No. PIB2010US-00153 and No. FIS2010-20979-C02-01.

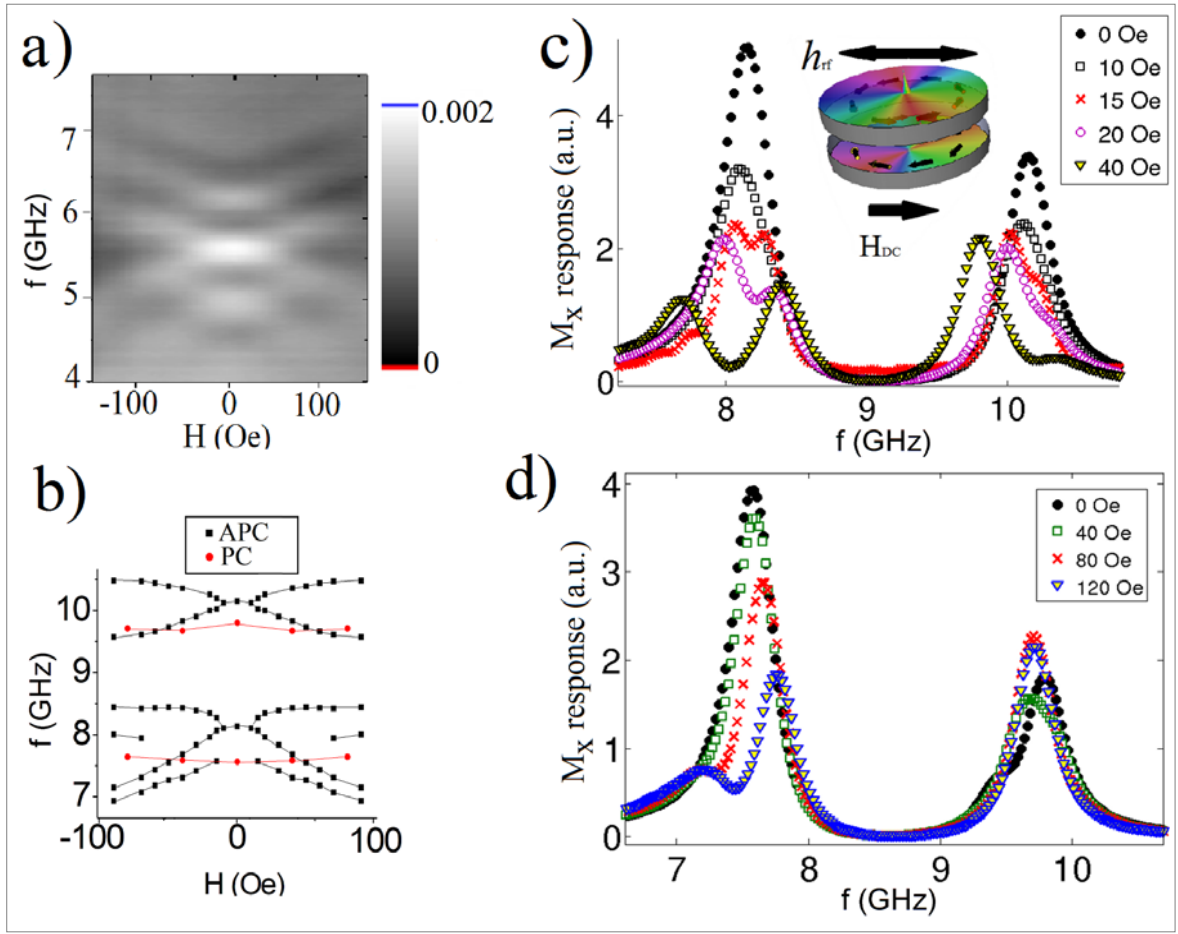

Fig. 1. Frequency excitation spectra of Py/Cu/Py nanopillars: a) measured broadband ferromagnetic resonance absorption spectra as a function of applied field. b) simulated eigenfrequencies as function of in-plane bias field for APC and PC states of the Py layers. c) simulated excitation spectra at low fields in APC state. d) simulated excitation spectra at low fields in PC state.

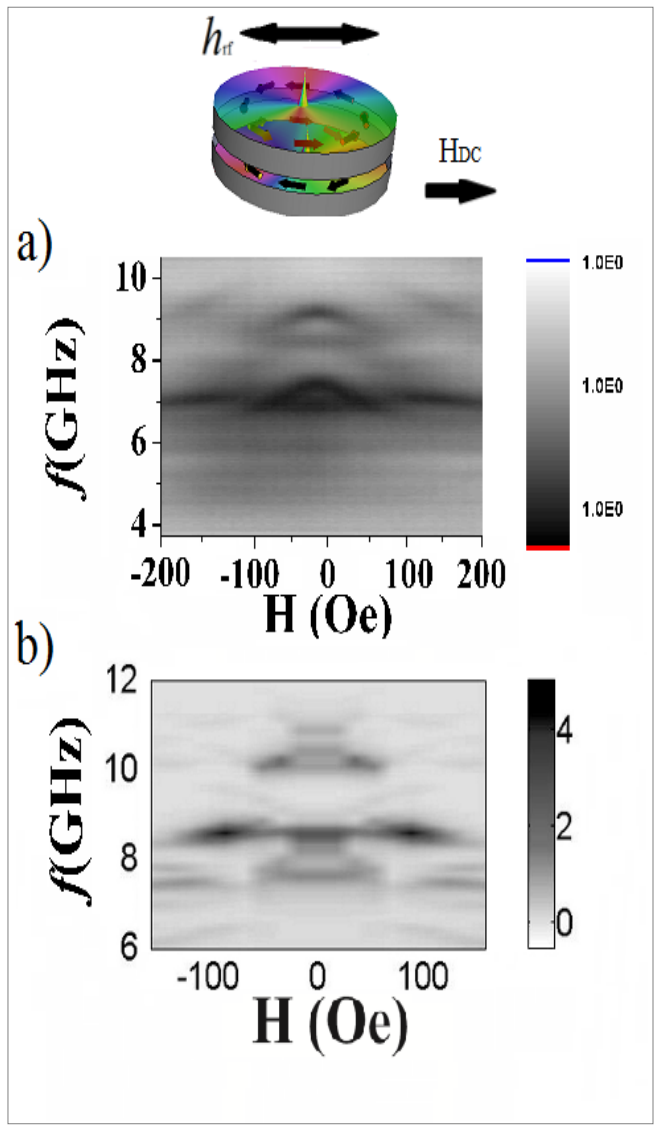

Fig.2. Spin wave excitation spectra of IEC Py/Cu(0.9)/Py nanopillars vs. in-plane bias field: (a) experimental broadband ferromagnetic resonance measurements and (b) dynamic micromagnetic simulations with 70% contribution from the APC nanopillars.

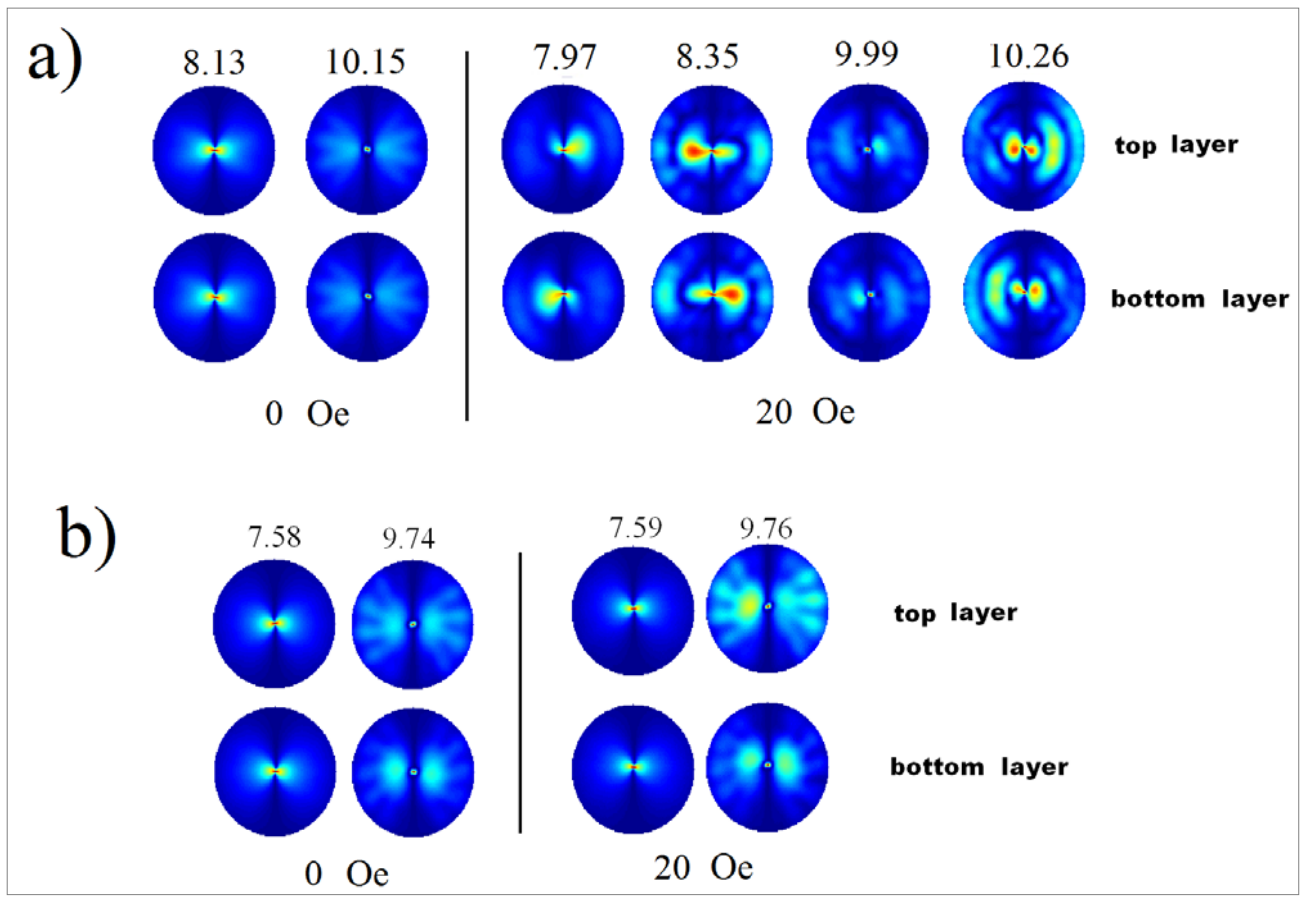

Fig. 3. Comparison of the simulated spin wave eigenmodes in the top and bottom layer of the tri-layer Py/Cu/Py nanopillars in (a) APC and (b) PC states (reduced dynamic magnetization component $\Delta M_x/M_S$ is presented). On the top of each mode image the corresponding frequency is specified (in GHz). When bias field is applied, there is 180º - phase shift of the moving mode profiles in the APC state, meanwhile in the PC state there is a mirror-like reflection of the phase along a central vertical line.